\documentclass[twocolumn,secnumarabic,amssymb, nobibnotes, aps, pre, floatfix]{revtex4-1}
\usepackage{amsmath}
\usepackage{wrapfig}
\usepackage{graphicx}
\usepackage{color}

\renewcommand{\vec}[1]{\ensuremath\boldsymbol{#1}}

\newcommand{\gphi}{\nabla\phi}

\begin{document}

\begin{abstract}
  A numerical and systematic parameter study of three-dimensional vesicle electrohydrodynamics is presented to investigate the effects of different fluid and membrane properties.
  The dynamics of vesicles in the presence of DC electric fields is considered, both in the presence and absence of linear shear flow.
  For suspended vesicles it is shown that the conductivity ratio and viscosity ratio between the interior and exterior fluids,
  as well as the vesicle membrane capacitance, substantially affect the minimum electric field strength required to induce a full Prolate-Oblate-Prolate transition.
  In addition, there exists a critical electric field strength above which a vesicle will no longer tumble when exposed to linear shear flow.
  
\end{abstract}

\title{Dynamics of Three-Dimensional Vesicles in DC Electric fields}%

\author{Ebrahim M. Kolahdouz and David Salac}%
\email[Corresponding author: ]{davidsal@buffalo.edu}
\affiliation{318 Jarvis Hall, Buffalo, NY 14226}
\date{February 29, 2015}%
\maketitle

\section{Introduction}
\label{sec:1.0}
	The interaction of lipid vesicles with external electric fields has been a subject of growing interest from both an application and a mechanistic 
	standpoint \cite{zimmermann1996electromanipulation,tresset2004microfluidic,dimova2007,vlahovska2009electrohydrodynamic}.
	One of the most striking phenomena in vesicle electrohydrodynamics is the Prolate-Oblate-Prolate (POP) transition \cite{schwalbe2011,mcconnell2013, salipante2014}.
	Consider a prolate vesicle where the major axis is initially aligned with the applied electric field. If the fluid encapsulated by the vesicle has a lower
	conductivity than the surrounding fluid there is a chance that the vesicle will transition to an oblate shape, where the major axes are perpendicular to the electric field.
	Over time the vesicle will evolve back into the prolate shape (aligned with the electric field) and stays in that equilibrium condition permanently \cite{salipante2014, schwalbe2011}.
	In order to observe the full POP transition the external electric field has to be strong enough to overcome the forces of the 
	vesicle membrane and the hydrodynamic forces of the fluid. 
	
	In physical experiments 
	capturing a full POP transition with a single DC pulse
	is not a trivial task \cite{salipante2014,Salipante2013thesis}.
	Strong fields can easily cause membrane poration which causes the vesicle to stay in the oblate shape permanently or to collapse before reaching the final prolate condition.
	Therefore, applying an electric field with the appropriate strength and duration is crucial in directing the behavior of the vesicle into the desired dynamics and configuration.
	
	Although a considerable number of experimental works in the past decade have been reported on vesicles exposed to electric fields \cite{dimova2007, salipante2014,Riske2005},
	a thorough numerical analysis of vesicle electrohydrodynamics is quite scarce. 
	Existing analytical studies are solely limited to nearly spherical vesicles \cite{vlahovska2009electrohydrodynamic, schwalbe2011} or
	planar membranes \cite{schwalbe2011lipid,seiwert2012stability} and make the Stokes assumption for the fluid flow.
	In this paper a systematic parameter study is carried out for a three-dimensional vesicle in the presence of DC electric fields.
	A full Navier-Stokes solver is employed which is able to capture possible inertia effects in the vesicle dynamics arising from the higher Reynolds number of electrohydrodynamic problems
	compared to typical hydrodynamic simulations in the absence of electric fields.
	The effects of fluid and vesicles properties on the electrohydrodynamic behavior are examined and the critical parameter thresholds required to observe a full POP transition are determined.
	This includes the effect of membrane capacitance and of the viscosity and conductivity ratios of the embedded and surrounding fluids for a range of different reduced volumes.
	Vesicles in the presence of combined shear flow and DC electric fields are also investigated and the critical field strength required to change a normally 
	tumbling vesicle into a tank-treading one is determined for different viscosity ratios.

\section{The mathematical model and numerical methods}
\label{sec:2.0}
	The model used in this paper is based on the computational approach presented by the authors in Ref. \cite{kolahdouzEHD2014}.  
	Consider a three-dimensional vesicle suspended in an aqueous solution where the inner and outer fluids may have differing 
	fluid viscosity and conductivity.
	The membrane is assumed to be inextensible with constant enclosed volume and surface area.
	The sum of the membrane forces is given as
	\begin{equation}
		-\kappa_c(\frac{H^3}{2}-2HK + \nabla^2_{s}H)\vec{n} + \gamma H\vec{n} -\nabla_s\gamma,
	\end{equation}
	where $\vec{n}$ is the 
	outward facing normal vector on the interface, $\gamma$ is the interface tension, $\kappa_c$ is the bending rigidity, 
	$H$ is the total curvature (sum of principle curvatures) and $K$ is the Gaussian curvature at the interface. 
	For a vesicle in the presence of an external electric field these forces are 
	balanced by the sum of the hydrodynamic and electric stresses.
	The hydrodynamic stress at the interface is given as 
	\begin{equation}
		\vec{\tau}_{hd}=\vec{n}\cdot[\vec{T}_{hd}]=\vec{n}\cdot\bigl(\vec{T}^+_{hd}-\vec{T}^-_{hd}\bigl),
	\end{equation}
	where $\vec{T}_{hd}=-p\vec{I}+\mu(\nabla\vec{u}+\nabla^T\vec{u})$ is the 
	bulk hydrodynamic stress tensor
	while $\mu$, $\vec{u}$ and $p$ refer to viscosity, fluid velocity, and pressure, respectively. 
	The $+$ sign is used when the interface is approached from the outer fluid while the $-$ sign 
	represents the interface being approached from the inner fluid. Square brackets, $[\;]$, indicate the jump
	of a quantity across the interface.
	
	The electric field stress at the interface is expressed as 
	\begin{equation}
		\vec{\tau}_{el}=\vec{n}\cdot[\vec{T}_{el}]=\vec{n}\cdot\bigl(\vec{T}^+_{el}-\vec{T}^-_{el}\bigl),
	\end{equation}
	with the Maxwell tensor, $\vec{T}_{el}$, given as 
	$\vec{T}_{el}=\epsilon(\vec{E}\vec{E} -\frac{1}{2}\left(\vec{E}\cdot\vec{E}\right)\vec{I})$,
	where $\vec{E}=-\nabla\Phi$ is the electric field, $\epsilon$ the permittivity, and $\Phi$ is the electric potential field.
	
	By employing a level set method to implicitly track the interface varying fluid properties at any point in the domain 
	can be written in a single relation, \textit{e.g.} 
	$\mu(\vec{x})=\mu^{-}+(\mu^{+}-\mu^{-})\mathcal{H}(\phi(\vec{x}))$ where $\mathcal{H}$ is the Heaviside function and $\phi$ is the level set function. 
	One is therefore able to write the momentum equations of binary fluids with varying properties into one single formulation. 
	
	Introduce $t_0$, $u_0$, $a$, and $E_0$ as the characteristic time, velocity, length, and electric field.
	Three important dimensionless quantities can then be defined as $Ca=\mu^{+}(1+\eta)a^3/(t_{0}\kappa_c)$, $Mn=t_0\epsilon^+ E_0^2/(\mu^+(1+\eta))$ and $Re=\rho u_0^2 t_0/\mu^+$.
	The quantity $Ca$ indicates the strength of the bending forces, while $Mn$ indicates the strength of the electric field effects and $Re$ is the Reynolds number.
	This allows for the single-fluid formulation of the Navier-Stokes equation to be written as \cite{kolahdouzEHD2014}
	\begin{equation}
	 \begin{split}
				\rho\frac{D \vec{u}}{Dt}=&-\nabla p + \frac{1}{Re}\nabla\cdot\left(\mu\left(\nabla\vec{u}+\nabla^{T}\vec{u}\right)\right)\\
					 & + \delta\left(\phi\right)\Arrowvert\gphi\Arrowvert\left(\nabla_s\gamma-\gamma H\gphi\right) \\
					 & + \frac{1}{Ca\;Re}\delta(\phi)\left(\frac{H^3}{2} -2KH+\nabla^2_s H \right)\gphi\\
					  & + \frac{Mn}{Re}\delta(\phi)\Arrowvert\gphi\Arrowvert\left[\epsilon\left(\vec{E}\vec{E}-\frac{1}{2}\left(\vec{E}\cdot\vec{E}\right)\vec{I}\right) \right]\cdot\vec{n}.
	  \end{split}
	  \label{eq:normalizedNS}
	\end{equation}
	Additionally, the fluid incompressibility and surface inextensibility constraints are given as $\nabla\cdot\vec{u}=0 $ and $\nabla_s\cdot\vec{u}=0$, respectively.
	Note that in Eq. (\ref{eq:normalizedNS}) all quantities are nondimensionalized and the
	singular contributions of the bending, tension, and electric field forces have been transformed into localized body force terms using 
	the level set and Dirac delta functions. All fluid properties are normalized with respect to the external fluid values.
	
	The leaky-dielectric model is used to obtain the electric potential in the domain.
	The electric potential is the solution to the Laplace equation in each region, $\nabla^2\Phi^{\pm}=0$, with an electric potential jump at the interface due to 
	the capacitive property of the membrane.
	This potential jump is called trans-membrane potential ($V_m$) and evolves over time by the non-dimensional relation, 
	\begin{equation}
		\bar{C}_m\frac{\partial V_m}{\partial t}+\vec{u}\cdot\nabla\left(\bar{C}_m V_m\right)=\vec{n}\cdot\vec{E^{+}},
		\label{eq:Vm_equation} 
	\end{equation}
	where  $\bar{C}_m=(C_m a)/(t_0 s^+)$ is the non-dimensional membrane capacitance, $C_m$ is the actual capacitance of the membrane and $\lambda=s^-/s^+$ is the conductivity ratio between the two fluids.
	Note that any trans-membrane conductance is currently ignored.
	In this work an immersed interface method is used to take into account the time-varying trans-membrane potential and the discontinuous fluid conductivities.
	This is done by deriving jump conditions for the electric potential and its first and second derivatives in order to achieve second order accuracy. 
	For details	of the numerical implementation interested readers are referred to Ref. \cite{kolahdouz2015}.

	For the electrohydrodynamic computations with no imposed shear flow the characteristic time, $t_0$, is set to the membrane charging time scale
	\textit{i.e.} $t_0=t_m=\frac{a C_m}{s^+} \left(\frac{1}{\lambda} + \frac{1}{2}\right)$. In simulations with 
	combined shear flow and an external electric field the time scale associated with the applied shear flow is used
	\textit{i.e.} $t_0=t_{\dot{\gamma}}=\dot{\gamma}_0^{-1}$ where $\dot{\gamma}_0$ is the applied shear rate. 
	Typical experimental values reported for physical properties are given as
	$a\approx 20\;\mu$m, $\kappa_c \approx 10^{-19}$ J, $s^+ \approx 10^{-3}$ S/m,,
	$\epsilon^+ = \epsilon^- \approx 10^{-9}$ F/m, $C_m \approx 10^{-2}$ F/m$^{2}$,
	$\rho \approx 10^3$ kg/m$^3$, $\dot{\gamma}_0 \approx 1$ $s^{-1}$, $\mu^-=\mu^+ \approx 10^{-3}$ Pa s and for the strong electric field $E_0=10^{5}$ V/m \cite{Riske2006,salipante2014,sadik2011vesicle}.	
	Using these values the dimensionless parameters in the presence of an external electric field and in absence of shearing flow are found to be  $Re=0.19$, $Ca=3.8\times10^{4}$, $Mn=18$ and $\bar{C}_m=0.1$.
	If the vesicle under the combined effects of electric field and imposed shear flow is considered,
	then the corresponding dimensionless parameters are  $Re=10^{-3}$, $Ca=10$ and $\bar{C}_m=2\times 10^{-4}$.
	Note that unlike the cases with $\dot{\gamma}_0=0$ where strong electric fields are applied, in the combination of the two effects only weak electric fields are considered therefore, the 
	dimensionless strength of the electric field is  within
	the range of $Mn=10$.
	
	The moving interface is modeled through
	the use of a semi-implicit, gradient-augmented level set jet scheme which is an extension of the original jet scheme introduced in \cite{Seibold2012}.
	The semi-implicit algorithm is implemented in three steps.
	The first step is the Lagrangian advection of the cell centers. In the second step the values at the cell centers are smoothed using a semi-implicit scheme. 
	This is done based on introducing a smoothing operator similar
	to the idea used in Refs. \cite{Kolahdouz20132, Smereka2003}.
	In the third step the center values are projected back to grid points and the jet of the solution including the gradients are updated; see Ref. \cite{kolahdouzEHD2014} for implementation details.
	
	To tackle the hydrodynamic problem a recently developed Navier-Stokes projection method is used.
	Both local and global fluid incompressibility and surface inextensibility conditions are satisfied in the numerical method 
	by splitting the pressure and tension correction terms into spatially-constant and spatially-varying components 
	and solving for the system of unknowns through the use of a Schur complement decomposition strategy.
	For details of the numerical approach and the special discretization see Ref. \cite{kolahdouzEHD2014}.

\section{Results}
\label{sec:3.0}

	In all the present simulations the vesicle surface area is fixed to $4\pi$ and the initial shape is a prolate ellipsoidal vesicle. The computational domain is
	a box with the size of $[-4.5,4.5]^3$.
	Periodic boundary conditions are taken in the $x-$ and $z-$directions while wall-boundary-conditions are taken in the $y-$direction.
	The electric field is imposed by setting an electric potential equal to $\Phi=-y$ on the wall-boundaries.
	In the case where shear flow is also applied the velocity at the wall-boundaries is taken to be $u=y$. Note that the electric field and fluid velocity
	have been normalized by $E_0$ and $u_0$.
	A Cartesian collocated mesh with uniform grid spacing is used and
	the grid spacing is chosen to be $h=0.075$ with a minimum time-step of $\Delta t=5\times10^{-4}$. Further information about the domain size, time step, and 
	grid spacing choices can be found in Ref. \cite{kolahdouzEHD2014}.
	
\begin{figure}[ht!]
		\centering
		\includegraphics[width=0.45\textwidth]{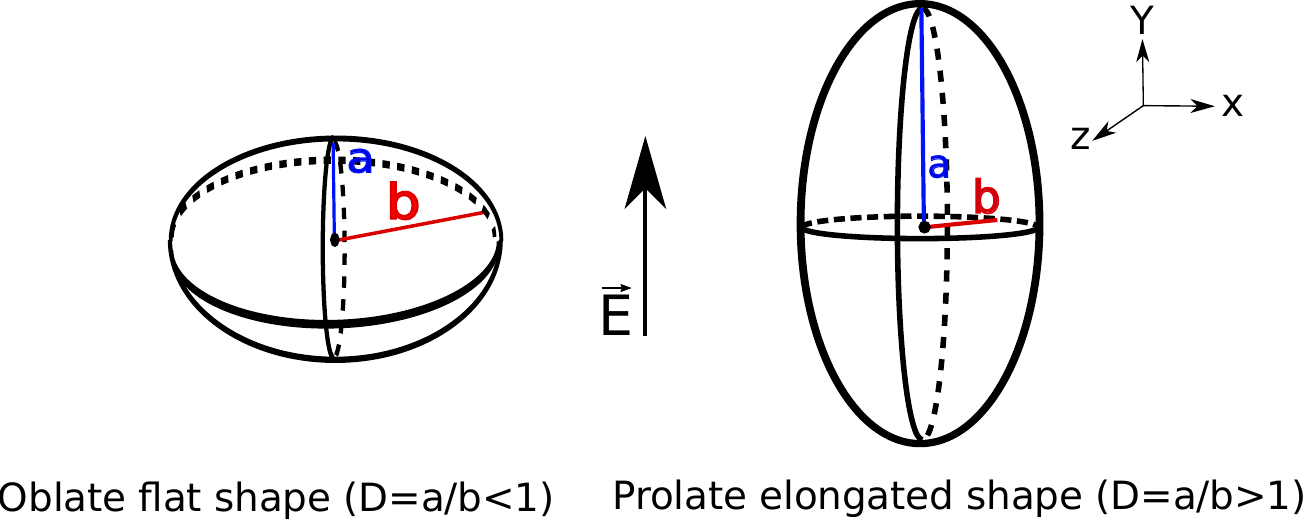}
		\caption{Sketch of the prolate and oblate shapes and their orientations with respect to the external electric field. 
		}
		\label{fig:axes}
	\end{figure}

	The vesicle configuration is described by the deformation parameter:
	\begin{align}
		D =& \frac{a}{b},	
	\end{align}
	where $a$ is the axis length of the vesicle in the direction parallel to the applied electric field and $b$ is the length of the axes
	perpendicular to the electric field, see Fig. \ref{fig:axes}. A prolate shape is given by $D>1$ (major axis aligned with the direction of the electric field)
	while the oblate shape has $D<1$ (major axis perpendicular to the direction of the electric field).
	The deformation, $D$, is directly calculated 
	from the interface location as described by the level set.

	The electrical properties of the two fluids, including the permittivity and electric conductivity, are among the most important factors in determining the dynamics
	of vesicles exposed to external DC electric fields. The ratio of the electric permittivity to the electric conductivity in each fluid gives the bulk charge relaxation time
	expressed as $t_c^{\pm}= \frac{\epsilon^{\pm}}{s^{\pm}}$. This time represents the rate at which each fluid can supply charges to the interface which is different for fluids with
	different electrical properties.
	The ratio of the charge relaxation time of the interior fluid to the exterior fluid is given by $t_c^-/t_c^+=\zeta/\lambda$ where $\zeta=\epsilon^{-}/\epsilon^{+}$
	and $\lambda=s^-/s^+$. This measure is of major importance in vesicle
	electrohydrodynamics and shows the competition between the two fluids in conducting charges towards the membrane. For
	typical vesicle experiments it is reasonable to assume $\zeta=1$ and hence the deviation of the conductivity ratio from one is the factor that determines different vesicle dynamics.
	
\begin{figure}[ht!]
	\centering
	\includegraphics[width=0.48\textwidth]{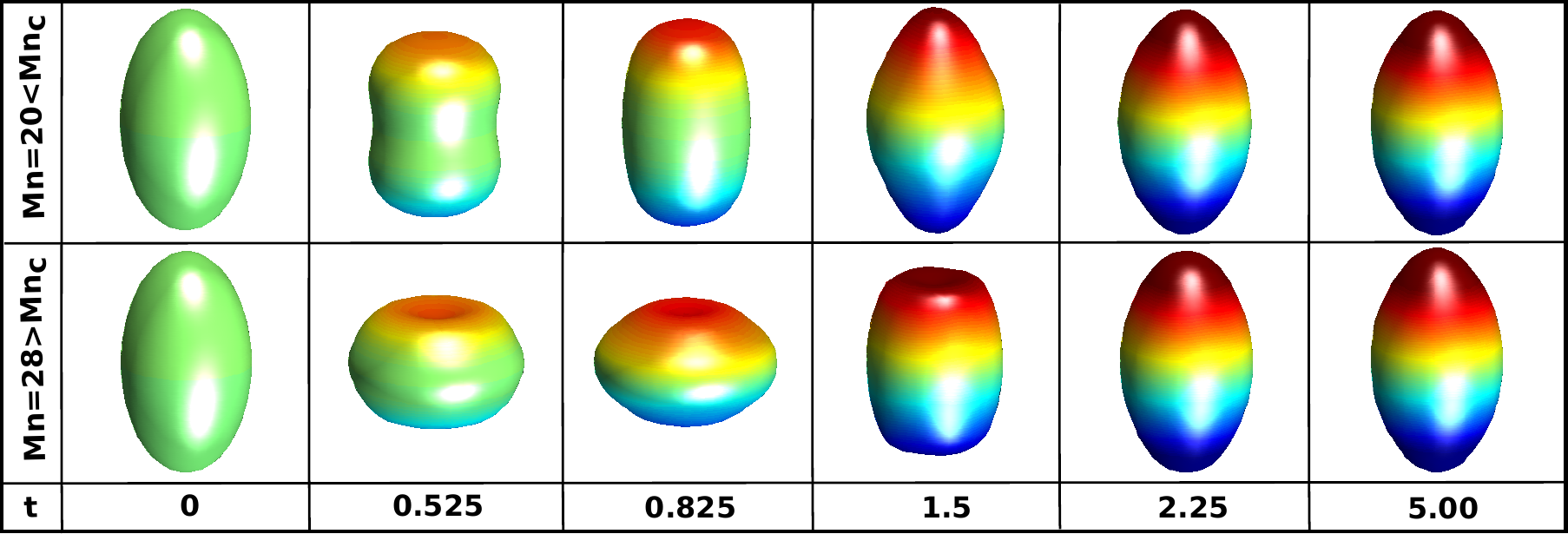}
	\caption{Simulation results showing the full POP transition (bottom series) vs. a vesicle that remains in the prolate regime (top series).
		The reduced volume in both simulations is $v=0.93$ and other simulation parameters are set to
		$Re=0.19$, $Ca=3.8\times10^{4}$, $\bar{C}_m=0.1$, $\lambda=0.1$ and $\eta=1.0$.		
		The colors on the surface (online version) indicate the distribution of the trans-membrane potential over time, 
		with darkest blue (lower pole) indicating 
		a membrane potential of $V_m=-1.6$ and darkest red (upper pole) indicating a membrane potential of $V_m=1.6$ for the saturated membrane capacitance.}
	\label{fig:POP_P}
	\end{figure}

	To illustrate the types of deformations in the full POP transition and how it compares to an incomplete transition
	the shape of a vesicle with two different electric field strengths (represented by $Mn$ values)
	are presented in Fig. \ref{fig:POP_P}.
	In the simulation results in the bottom series a strong electric field of $Mn=28$ is applied while a weaker field, $Mn=20$, is used for the results in the top row.
	In the former case, the electric forces are strong enough to overcome the membrane forces of the vesicle and the viscous forces of the fluid and this
	drives the vesicle into the oblate state where $D<1$. However, when the weak electric field is applied the initial prolate vesicle 
	transitions into to a slightly less prolate profile but is unable to obtain the prolate to oblate transition. 
	The final shape is the same in both cases.
	This indicates that there exists a critical electric field strength, $Mn_{c}$, above which a full POP transition occurs to the suspended vesicle exposed to an external DC field. 
	As will be shown below, this critical electric field strength depends on a number of material parameters.

\begin{figure}[ht!]
		\centering
		\includegraphics[width=0.4\textwidth]{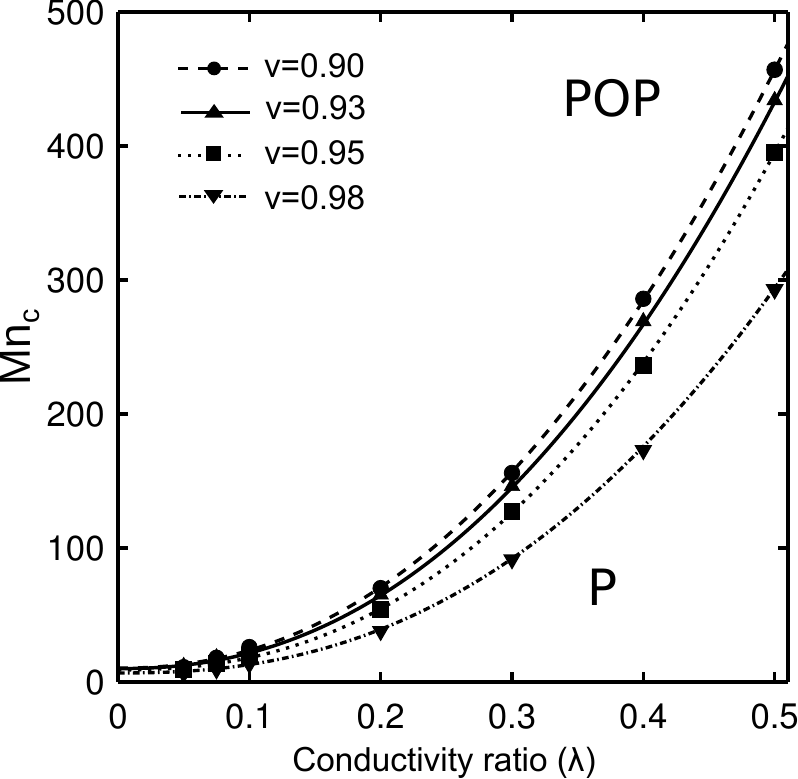}
		\caption{The effect of the conductivity ratio, $\lambda$, on the POP transition. The region above the curve shows where the POP transition happens while the region below the curve
		represents a vesicle that remains prolate for all the time. The transition is plotted for four different reduced volumes, $v=0.90,0.93,0.95,$ and  $0.98$.
		The two regions are separated by the critical field strength, $Mn_c$ where the dynamics changes from purely prolate to prolate-oblate-prolate. The nondimensional parameters are
		   set to $Re=0.19$, $Ca=3.8\times10^{4}$, $\eta=1.0$ and $\bar{C}_m=0.1$.}
		\label{fig:CondMnPOP}
	\end{figure}

	To study the role of the fluid conductivity ratio in POP dynamics the critical Mason number, $Mn_c$, is determined for various 
	conductivity ratios.
	This behavior is illustrated in Fig. \ref{fig:CondMnPOP} for a range of different reduced volumes. 
	The region above each curve shows where the POP transition occurs while the region below the curves indicates the $Mn$ values that
	are not strong enough to force the vesicle to full POP transition.
	As the the conductivity ratio $\lambda$ increases, larger and larger electric field strengths are required to induce the POP transition. 
	Theoretically speaking, for a vesicle with $\lambda=1$, \textit{i.e.} matched conductivity between the two fluids, the $Mn_c$ approaches infinity. In fact, as will
	be demonstrated in Sec. \ref{sec:3.0} it is not possible for a vesicle with match conductivity to achieve a POP transition.
	
	The membrane capacitance is another important factor in determining the dynamics of the vesicle in a DC field. In particular, the capacitance appears as a primary parameter
	in the time scale associated with the membrane charging given, $t_m$. The larger the capacitance the larger duration of time electric field forces
	can act on the membrane. In the context of the POP transition an interesting 
	comparison can be made between the membrane charging 
	timescale and the electrohydrodynamic time scale of the applied field given by $t_{ehd}=\frac{\mu^+(1+\eta)}{\epsilon^+ E_0^2}$. The electrohydrodynamic time scale largely
	depends on the electric field strength: a strong electric field is associated with a shorter electrohydrodynamic time. 
	In theory, for a nearly spherical vesicle a full POP transition will occur if $t_{ehd}<t_m$ \cite{salipante2014}.
	In such situations electric field forces have sufficient time to induce a full POP deformation before the trans-membrane potential is saturated at time $t_m$. 
	As the capacitance increases the charging time of the membrane capacitance, $t_m$, increases as well which implies that even with a weaker electric field (larger $t_{ehd}$) 
	a full POP transition is still possible
	as long as $t_{ehd}$ is sufficiently smaller than $t_m$. 
	
	Note that the membrane capacitance affects the amount of deformation as well. An increase of $t_m$ due to larger capacitance leads to larger deformations 
	as the electric forces only influence the dynamics of the membrane in proportion to the time that they are allowed to act.
	Since the capacitance is a major factor in determining this time the implication is that the
	deformations are smaller for smaller dimensionless capacitance, $\bar{C}_m$, and deformations grow larger
	as $\bar{C}_m$ does.
	This manifests the importance of the 
	membrane capacitance in the type and magnitude of deformations.

\begin{figure}[ht!]
	\centering
	\includegraphics[width=0.4\textwidth]{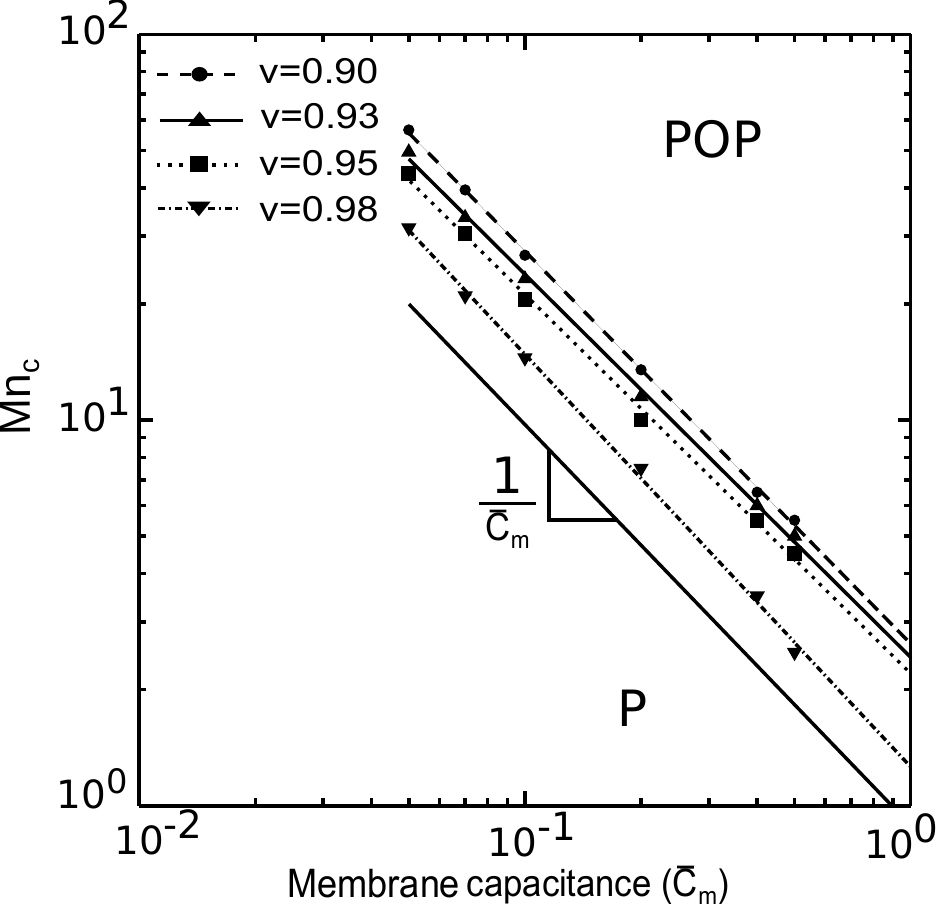}
	\caption{ The effect of the membrane capacitance, $\bar{C}_m$, on the POP transition. The region above the curve shows where the POP transition occurs while the region below the curve
		represents a vesicle that remains prolate for all the time. The transition is plotted for four different reduced volumes, $v=0.90,0.93,0.95$, and $0.98$.
		The two regions are separated by the critical field strength, $Mn_c$, where the dynamics change from purely prolate to prolate-oblate-prolate. 
		The nondimensional parameters are set to $Re=0.19$, $Ca=3.8\times10^{4}$, $\eta=1.0$ and $\lambda=0.1$.}
	\label{fig:CmMnPOP}
	\end{figure}

	Figure \ref{fig:CmMnPOP} shows how changes in the membrane capacitance leads to a vastly different critical field strengths, $Mn_c$, needed to observe a full POP transition. This
	behavior is consistent across all investigated vesicle reduced volumes. For the four reduced volumes investigated the critical field strength scales as the inverse of
	the membrane capacitance, $Mn_c \propto \bar{C}_m^{-1}$. This relationship will be further discussed in Sec. \ref{sec:Discussion}.
	

\begin{figure}[ht!]
	\centering
	\includegraphics[width=0.4\textwidth]{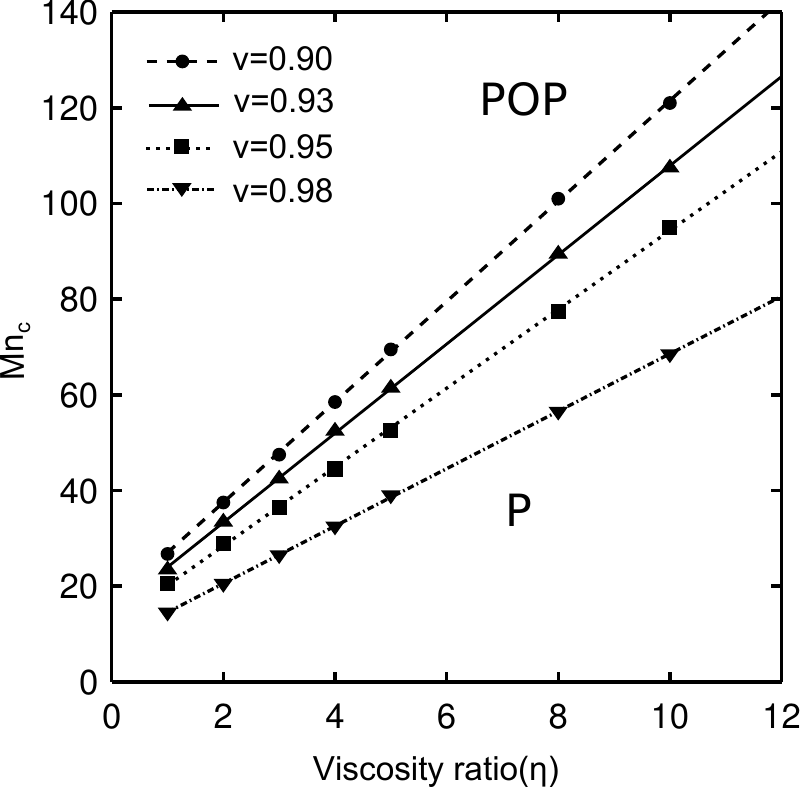}
	\caption{The viscosity ratio effect on the vesicle behavior transition from the prolate regime (region below the curve) to POP regime (region above the curve).
		The two regions are separated by the critical field strength, $Mn_c$ where the dynamics changes from one to another.
		This transition is plotted for four different reduced volumes, $v=0.90$, 0.93, 0.95 and 0.98. The non-dimensional parameters are
		set to $Re=0.19$, $Ca=3.8\times10^{4}$, $\bar{C}_m=0.1$ and $\lambda=0.1$.}
	\label{fig:ViscMnPOP}
	\end{figure}

	The  effect of the viscosity ratio ($\eta=\mu^-/\mu^+$) on the dynamics of the vesicle is shown in Fig. \ref{fig:ViscMnPOP} for four different reduced volumes. 
	Numerical experiments demonstrate that as $\eta$ increases a stronger electric field is needed to overcome the resisting hydrodynamic forces of the fluids.
	A larger viscosity ratio is associated with a larger viscous damping force which in turn leads to slower dynamics and thus a larger $Mn_c$  required to induce the full POP transition.
	The critical electric field strength appears to be linearly related to the viscosity, $Mn_c \propto \eta$.
	
	One consequence of these results is that it appears that given the same electrical properties, 
	smaller reduced volumes require a stronger field to achieve a full POP transition. This is most likely dependent on the initial shape,
	as smaller reduced volumes begin as more prolate (more stretched out) vesicles. This means that as the reduced volume decreases
	larger deformations are required to obtain the prolate-oblate transition. Future work will investigate this, as it is possible to have many different
	initial vesicle shape conditions for the same reduced volume.
	
	Next, the combined effects of an electric field and shear flow are considered. 
	In this case the strength of the external electric field is increased from zero to investigate 
	the changes in the dynamics of the vesicle.
	Note that here the shear flow time scale is used as the characteristic time-scale. With typical 
	experimental values this results in non-dimensional parameters $Re=0.001$, $Ca=10$ and  $\bar{C}_m=2\times10^{-4}$.
	For typical situations the shear flow time scale is approximately $t_{\dot{\gamma}}=1 s$
	while the membrane charging time scale is approximately $10^{-3}$ s \cite{kolahdouzEHD2014}. Therefore, in this situation the trans-membrane potential is 
	taken to respond instantly to changes in the membrane configuration and is thus at a pseudo-steady-state.
	
	The effect of the electric field is to align the vesicle vertically with the direction of the field. 
	Therefore, for a vesicle in the tumbling regime an increase in $Mn$ will result in a slower tumbling period. A close observation of vesicle dynamics under the combined 
	shear and an external weak electric field reveals that 
	the periodic behavior in this situation is more akin to vacillating-breathing motion where the vesicle does not follow a rigid-body-like rotation anymore, but instead 
	the poles retract and additional deformations are induced in the shear plane. 
	If the field is strong enough, the electric field forces acting against the
	shearing forces of the fluid can eventually bring the vesicle into the tank-treading regime.

	\begin{figure}[ht!]
	\centering
	\includegraphics[width=0.48\textwidth]{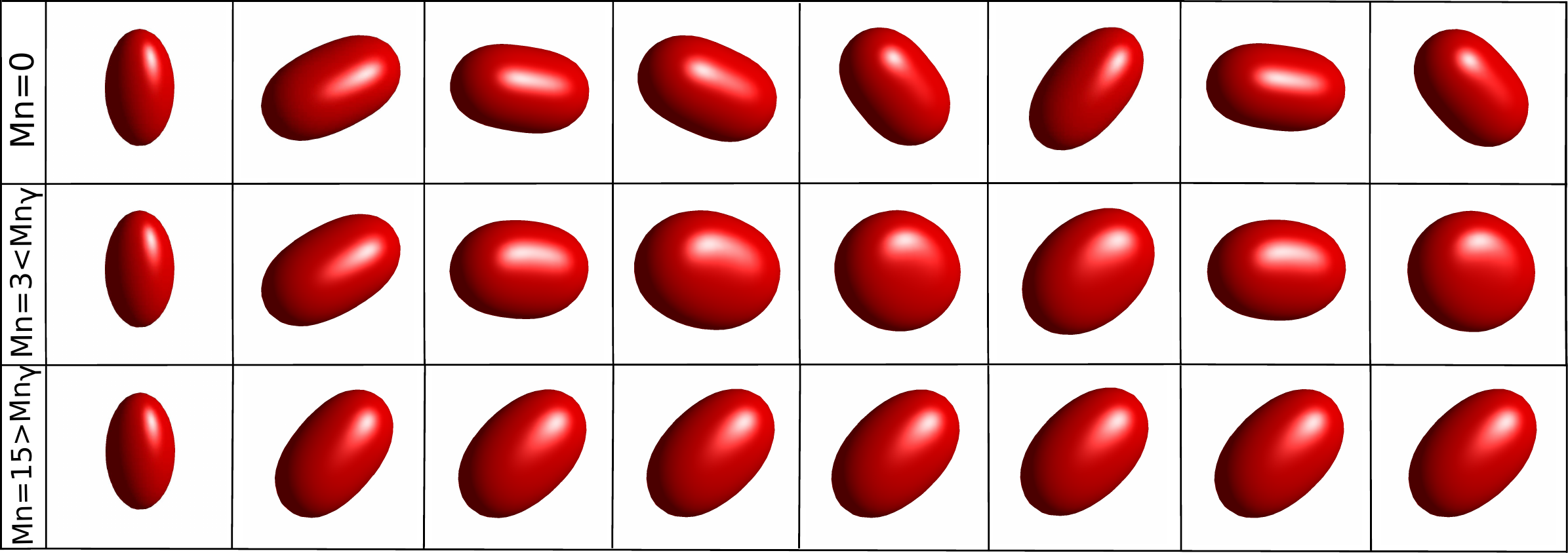}
	\caption{Dynamics of a vesicle under the combined effects of shearing flow and an applied electric field at various times.
		The reduced volume is $v=0.93$ and a viscosity ratio of $\eta=10$ is used. 
		Other simulation parameters are $Re=10^{-3}$, $Ca=10$ and  $\bar{C}_m=2\times10^{-4}$.
		From left to right the snapshots are given at $t=0.0, 2.5, 6.0, 7.0, 8.0, 10.0, 15.0, 17.0$. The top row shows the
		tumbling vesicle in the absence of external electric field ($Mn=0$). 
		The middle row corresponds to a vesicle with a vacillating-breathing-like motion under an electric field strength of $Mn=3$.
		The bottom row shows a tank-treading vesicle under an strong electric field strength of $Mn=15$. }
	\label{fig:TBTTMn}
	\end{figure}
		
	These different dynamic regimes are shown in Fig. \ref{fig:TBTTMn}.
	In this example the vesicle has a reduced volume of $v=0.93$ and the viscosity ratio is $\eta=10$. 
	This ratio is larger than the critical viscosity ratio needed to obtain tumbling in shear flow when in the absence of an electric field.
	The top row in the figure shows the dynamics over time when the electric field is not present in the simulation ($Mn=0$).
	As would be expected the vesicle tumbles with
	the profile of the vesicle remaining nearly constant and rotating periodically in time.
	The vacillating-breathing-like motion is seen for the results shown in the middle row. In this simulation a weak electric field of $Mn=3$ is used.
	Unlike the tumbling motion, in this regime the profile of the vesicle undergoes changes over time as it periodically rotates about the vorticity axis.
	Finally the bottom row demonstrates the vesicle dynamics of the same vesicle under an electric field with a stronger strength ($Mn=15$). 
	The vesicle reaches an equilibrium, tank-treading inclination angle and stays in that position permanently.

\begin{figure}[ht!]
	\centering
	\includegraphics[width=0.45\textwidth]{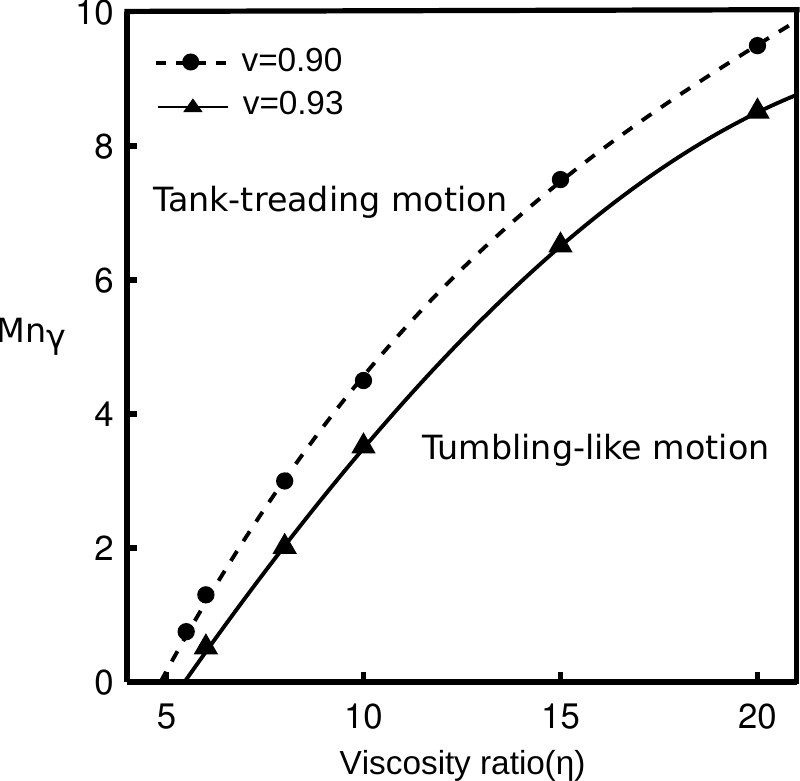}
	\caption{Tumbling-like to tank-treading transition for vesicles with two different reduced volumes, $v=0.90$ and $v=0.93$, under the combined effect of imposed shear 
		flow and weak DC electric fields.
		The non-dimensional parameters are set to $Re=0.001$, $Ca=10$, $\chi=1$, $\bar{C}_m=2\times10^{-4}$ and $\lambda=0.1$. 
		Membranes with $Mn \neq 0$ have a pseudo-steady-state trans-membrane potential. The region above the curve
		is the tank-treading regime while the region below the curve shows the tumbling-like regime. Increasing the field strength from zero 
		gives rise to a damping force against the motion of a normally tumbling vesicle.}
	\label{fig:TBTTEHD}
	\end{figure}

	The critical Mason number at which the transition from tumbling-like motion (either tumbling or vacillating-breathing) to tank-treading occurs is denoted as $Mn_{\gamma}$. 
	The required $Mn_{\gamma}$ for two different reduced volumes over a range of viscosity ratios is plotted in Fig. \ref{fig:TBTTEHD}.  
	
\section{Discussion}
\label{sec:Discussion}
	To better understand why a vesicle will undergo a prolate-oblate-prolate transition in the presence of an electric field
	let us consider the electric field forces acting on the membrane. For simplicity and to match 
	the computational results assume that the electric permittivity ratio is one, $\epsilon^+=\epsilon^-$.
	Next, consider only the normal electric field forces on the interface. Tangential electric field forces
	will drag the fluid into motion but do not induced changes in the vesicle membrane.
	
	It is straightforward to write the normal electric force in terms of the normal and tangential components of the electric field,
	\begin{equation}
		\vec{\tau}_{el}\cdot\vec{n}=\frac{\epsilon^{+}}{2}\bigl( ({E_n^+}^{2} - {E_n^-}^{2}) - ({E_t^+}^{2} + {E_b^+}^{2} - {E_t^-}^{2} - {E_b^-}^2 ) \bigl),
		\label{eq:normal_Eforce}
	\end{equation}
	where $E_t$ and $E_b$ are the tangential and $E_n$ is the normal electric field at that point. Note that the notation here is ${E_n^+}^{2}=\left({E_n^+}\right)^{2}$
	and ${E_n^-}^{2}=\left({E_n^-}\right)^{2}$.
	
\begin{figure}[ht!]
		\includegraphics[width=0.48\textwidth]{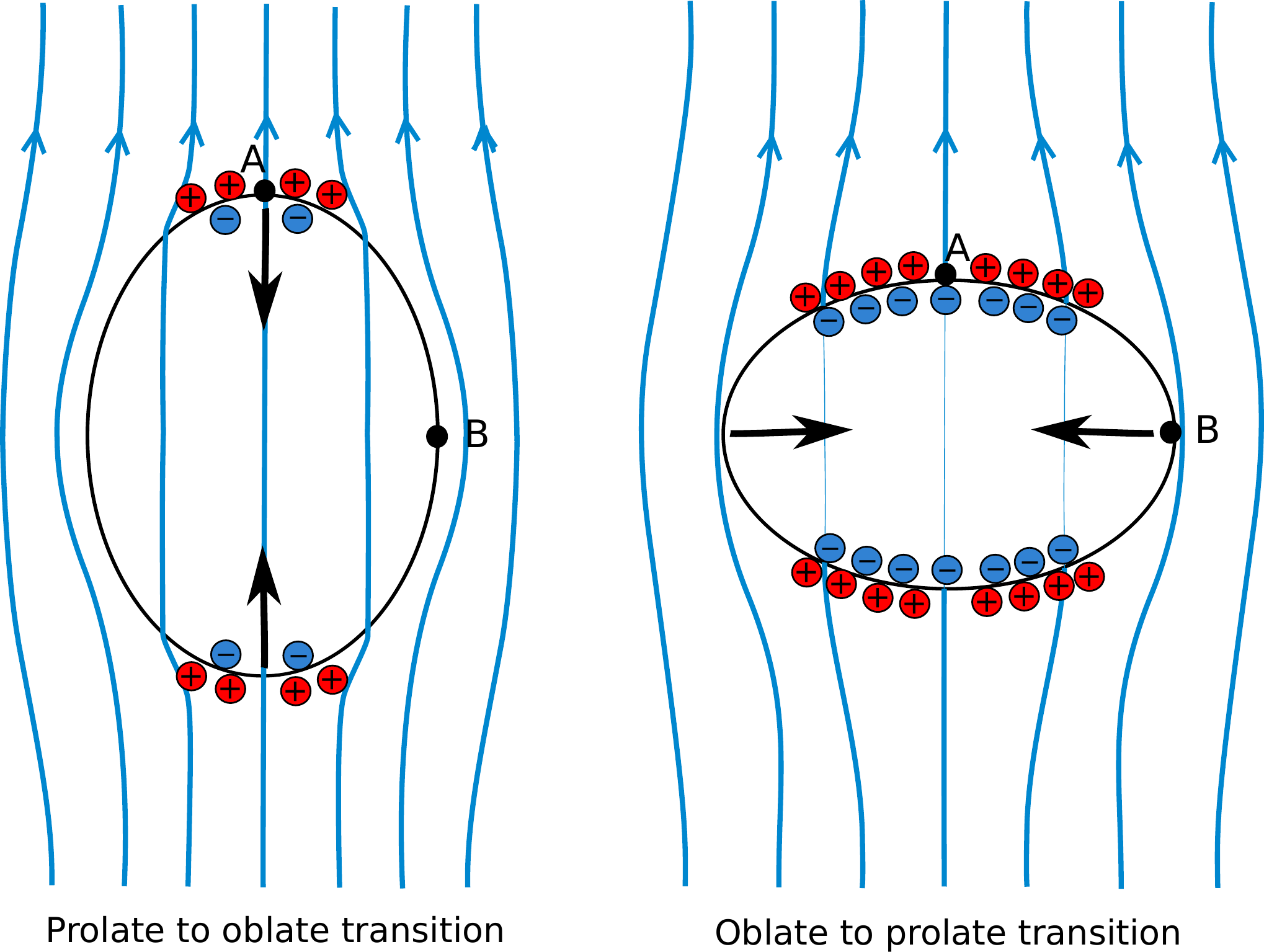}
		\caption{A sketch of the electric field, induced charges and electric forces when an initially prolate vesicle is transitioning to the oblate shape (left figure) and when the
			oblate vesicle is evolving back to the equilibrium prolate transition (right figure).}
		\label{fig:Normal_Eforces}
	\end{figure}

	This clearly shows that both normal and tangential electric field components contribution to membrane deformation. 
	To further simplify the discussion, consider the two points $A$ and $B$ located respectively at the top pole and the equator of the vesicle, see Fig. \ref{fig:Normal_Eforces}.
	Due to their particular locations with regard to the external field the normal electric field force at these two points is simplified to
	\begin{align}
		\vec{\tau}^A_{el}\cdot\vec{n}&=\frac{\epsilon^{+}}{2}\bigl( {E_n^+}^{2} - {E_n^-}^{2} \bigl), \label{eq:elecforceA_1}\\ 
		\vec{\tau}^B_{el}\cdot\vec{n}&=\frac{-\epsilon^{+}}{2}\bigl({E_t^+}^{2} + {E_b^+}^{2} - {E_t^-}^{2} - {E_b^-}^2  \bigl).   \label{eq:elecforceB}
	\end{align}
	One of the conditions on the interface is the continuity of the normal electric field, $s^+ E_n^+-s^- E_n^-=0$. It is therefore possible 
	to write $E_n^+=\lambda E_n^-$ and thus at point $A$ the normal electric force becomes
	\begin{equation}
		\vec{\tau}^A_{el}\cdot\vec{n}=\frac{\epsilon^+(\lambda-1)}{2}{E_n^-}^2.
		\label{eq:elecforceA_2}
	\end{equation}
	
	After the electric field is applied charges from both the enclosed and surrounding fluids start migrating towards the interface.
	However, due to the differing electric conductivities the charges accumulate at differing rates.
	This leads to an apparent charge, 
	\begin{equation}
		Q=\epsilon^{+}(E_n^+ - E_n^-),
	\end{equation}
	assuming matching permittivities.
	
	Let the electric potential be activated at time 0. The electric fields inside and outside the vesicle are initially matched and non-zero: 
	$E_n^+ = E_n^-\neq 0$, $E_t^+ = E_t^-\neq 0$, and $E_b^+ = E_b^-\neq 0$. According to Eq. (\ref{eq:elecforceB}) the force along the 
	equator is zero. The direction of the force at the vesicle poles depends on the conductivity ratio. When $\lambda<1$ the normal force
	is in the $-\vec{n}$ direction, which is a vertically compressive force. On the other hand, when $\lambda>1$ the force is extensional
	and thus the POP transition can not occur. This demonstrates the necessary condition of $\lambda<1$ when the fluid permittivities are matched.
	
	Now consider what occurs when $t\rightarrow \infty$. Given enough time charges are able to accumulate on both sides of the interface. Therefore $Q=0$
	and the electric field inside the vesicle becomes zero, $\vec{E}^-=\vec{0}$ \cite{salipante2014}.
	Accordingly, for matched permittivities $E_n^-=E_n^+=0$ while $E_t^-=0$ and $E_b^-=0$. Note that in this case $E_t^+\neq 0$ and $E_b^+\neq 0$.
	From Eq. (\ref{eq:elecforceA_1}) this results in a zero normal force at point $A$ while
	along the equator, point $B$, the normal force reduces to a purely compressive force. Notice that this behavior is independent of the fluid electrical 
	conductivities, indicating that the final shape will always be a prolate one.	
	
	Finally, a rational behind the influence of the three main parameters investigated here, membrane capacitance, fluid conductivity ratio, and 
	fluid viscosity ratio, on the critical electric field strength needed to obtain a POP transition is presented.
	First examine the contribution of the membrane capacitance. The membrane capacitance 
	influences how fast the trans-membrane potential reaches a steady state value, Eq. (\ref{eq:Vm_equation}). The time
	it takes for the trans-membrane potential to reach steady state is inversely proportional to $\bar{C}_m$. As the amount of time 
	that the electric forces can act on the membrane is directly related to the membrane-charging time, a decrease in $t_m$ 
	results in a system which does not have enough time to respond and obtain a full POP transition. Therefore the 
	Mason number must increase at a rate proportional to $\bar{C}_m^{-1}$ to ensure that the POP transition occurs.
	
	Next consider the viscosity ratio, $\eta$. It was stated in Sec. \ref{sec:3.0} that for a POP transition to occur the electrohydrodynamic time, $t_{ehd}$,
	needs to be smaller than the membrane charging time, $t_m$. 
	As the electrohydrodynamic time increases linearly with an increase in the viscosity ratio, $t_{ehd}\propto 1+\eta$, the membrane charging time must also increase linearly to $\eta$ 
	to maintain the $t_{ehd}<t_m$ relation. Recall that for results in the absence of shear flow the characteristic time
	was set to the membrane charging time, $t_0=t_m$, and therefore the Mason number depends linearly on $t_m$. Thus, an increase in the 
	viscosity ratio results in a linear increase in $t_{ehd}$, which results in a linear increase of $t_m$ so that the POP transition can occur, which
	finally leads to a linear increase of $Mn_c$.
	
	The conductivity ratio presents a more complicated picture. Based on the previous discussion only $\lambda<1$ will result in a POP transition
	and thus all cases with $\lambda\geq 1$ are ignored. Examining $t_m$ it is clear that an increase in the conductivity ratio will result in a decrease
	of the membrane-charging time. This is due to the fact that as $\lambda \rightarrow 1$ the fluids are able to supply charges to the interface at the same rate
	and thus a charge mismatch does not occur. Following the discussion regarding the viscosity ratio, a corresponding decrease in $t_{ehd}$ would
	be required for a POP transition to occur. This is achieved by increasing the electric field strength and therefore increasing $Mn$. Analysis
	of the equations results in a scaling of $Mn_c \propto (2 \lambda)/(2+\lambda)$, which can not account for the rapid increase of $Mn_c$ 
	as $\lambda$ approached 1, Fig. \ref{fig:CondMnPOP}. It is thus necessary to turn to the forces acting on 
	the membrane and how they relate to $\lambda$, Eq. (\ref{eq:elecforceA_2}). Clearly, as the conductivity ratio approaches 1 the forces acting on the membrane
	asymptotically approach zero. It is therefore expected that rapid growth of the electric field would be needed obtain a full POP transition. It is suspected
	that at small conductivity ratios the increase in $Mn_c$ is dominated by $(2 \lambda)/(2+\lambda)$ while as $\lambda \rightarrow 1$ the growth
	becomes dominated by $(1-\lambda)^{-1}$. Further investigations are needed to verify this.
	
	\section{Concluding remarks}
	\label{sec:4.0}
	In this paper a parameter space study of 3D vesicles exposed to DC electric fields was presented. 
	In particular, the focus of this work was on the effects of different fluid and membrane parameters 
	on the transition between prolate and oblate shapes for a suspended vesicle, in addition to the transition from the tumbling to tank-treading regime
	of a vesicle under the combined effects of a DC electric field and shear flow. 
	
	The electric field strength, indicated by the critical Mason number, $Mn_c$, required for a POP transition was determined.
	A decrease in the reduced volume of the vesicle was always accompanied with an increase in the critical Mason number.
	While an increase in the viscosity ratio between the inner and outer fluids brought about an almost linear increase of $Mn_c$, 
	the change in the dynamics of the vesicle 
	by varying the conductivity ratio between the two fluids and the membrane capacitance of the vesicle was found to be more dramatic and substantial. 
	
	The results of the vesicle behavior under the combined effects of shear flow and weak DC electric fields
	revealed the remarkable influence of the electric field in changing the standard behaviors of tank-treading and tumbling vesicles.
	Investigations showed that the application of the electric field
	generates a damping force against the motion of the normally tumbling vesicle. If this force is strong enough the tumbling
	vesicle stops the flipping tumbling behavior and undergoes a tank-treading motion 
	The critical $Mn_\gamma$ for this transition to occur was determined over a range of different viscosity ratios for vesicles
	with two different reduced volumes.
	
	The results and analysis presented here assumed matched fluid permittivities and only considered a single initial vesicle shape. It is 
	expected that different initial vesicle shapes could result in differing dynamics while the equilibrium shape would remain the same
	as presented here. Additionally, a relaxation of the matched permittivity condition could result in behavior not observed here. Future work
	will explore the influence of initial vesicle shape and permittivity differences on the dynamics of vesicles exposed to DC electric fields.
	
\begin{acknowledgments}
Authors acknowledge financial support by the U.S. National Science Foundation Grant No. 1253739.
\end{acknowledgments}


\begin{thebibliography}{18}%
\makeatletter
\providecommand \@ifxundefined [1]{%
 \@ifx{#1\undefined}
}%
\providecommand \@ifnum [1]{%
 \ifnum #1\expandafter \@firstoftwo
 \else \expandafter \@secondoftwo
 \fi
}%
\providecommand \@ifx [1]{%
 \ifx #1\expandafter \@firstoftwo
 \else \expandafter \@secondoftwo
 \fi
}%
\providecommand \natexlab [1]{#1}%
\providecommand \enquote  [1]{``#1''}%
\providecommand \bibnamefont  [1]{#1}%
\providecommand \bibfnamefont [1]{#1}%
\providecommand \citenamefont [1]{#1}%
\providecommand \href@noop [0]{\@secondoftwo}%
\providecommand \href [0]{\begingroup \@sanitize@url \@href}%
\providecommand \@href[1]{\@@startlink{#1}\@@href}%
\providecommand \@@href[1]{\endgroup#1\@@endlink}%
\providecommand \@sanitize@url [0]{\catcode `\\12\catcode `\$12\catcode
  `\&12\catcode `\#12\catcode `\^12\catcode `\_12\catcode `\%12\relax}%
\providecommand \@@startlink[1]{}%
\providecommand \@@endlink[0]{}%
\providecommand \url  [0]{\begingroup\@sanitize@url \@url }%
\providecommand \@url [1]{\endgroup\@href {#1}{\urlprefix }}%
\providecommand \urlprefix  [0]{URL }%
\providecommand \Eprint [0]{\href }%
\providecommand \doibase [0]{http://dx.doi.org/}%
\providecommand \selectlanguage [0]{\@gobble}%
\providecommand \bibinfo  [0]{\@secondoftwo}%
\providecommand \bibfield  [0]{\@secondoftwo}%
\providecommand \translation [1]{[#1]}%
\providecommand \BibitemOpen [0]{}%
\providecommand \bibitemStop [0]{}%
\providecommand \bibitemNoStop [0]{.\EOS\space}%
\providecommand \EOS [0]{\spacefactor3000\relax}%
\providecommand \BibitemShut  [1]{\csname bibitem#1\endcsname}%
\let\auto@bib@innerbib\@empty
\bibitem [{\citenamefont {Zimmermann}\ and\ \citenamefont
  {Neil}(1996)}]{zimmermann1996electromanipulation}%
  \BibitemOpen
  \bibfield  {author} {\bibinfo {author} {\bibfnamefont {U.}~\bibnamefont
  {Zimmermann}}\ and\ \bibinfo {author} {\bibfnamefont {G.~A.}\ \bibnamefont
  {Neil}},\ }\href@noop {} {\emph {\bibinfo {title} {{Electromanipulation of
  cells}}}}\ (\bibinfo  {publisher} {CRC press},\ \bibinfo {year}
  {1996})\BibitemShut {NoStop}%
\bibitem [{\citenamefont {Tresset}\ and\ \citenamefont
  {Takeuchi}(2004)}]{tresset2004microfluidic}%
  \BibitemOpen
  \bibfield  {author} {\bibinfo {author} {\bibfnamefont {G.}~\bibnamefont
  {Tresset}}\ and\ \bibinfo {author} {\bibfnamefont {S.}~\bibnamefont
  {Takeuchi}},\ }\href@noop {} {\bibfield  {journal} {\bibinfo  {journal}
  {Biomedical Microdevices}\ }\textbf {\bibinfo {volume} {6}},\ \bibinfo
  {pages} {213} (\bibinfo {year} {2004})}\BibitemShut {NoStop}%
\bibitem [{\citenamefont {Dimova}\ \emph {et~al.}(2007)\citenamefont {Dimova},
  \citenamefont {Riske}, \citenamefont {Aranda}, \citenamefont {Bezlyepkina},
  \citenamefont {Knorr},\ and\ \citenamefont {Lipowsky}}]{dimova2007}%
  \BibitemOpen
  \bibfield  {author} {\bibinfo {author} {\bibfnamefont {R.}~\bibnamefont
  {Dimova}}, \bibinfo {author} {\bibfnamefont {K.~A.}\ \bibnamefont {Riske}},
  \bibinfo {author} {\bibfnamefont {S.}~\bibnamefont {Aranda}}, \bibinfo
  {author} {\bibfnamefont {N.}~\bibnamefont {Bezlyepkina}}, \bibinfo {author}
  {\bibfnamefont {R.~L.}\ \bibnamefont {Knorr}}, \ and\ \bibinfo {author}
  {\bibfnamefont {R.}~\bibnamefont {Lipowsky}},\ }\href@noop {} {\bibfield
  {journal} {\bibinfo  {journal} {Soft matter}\ }\textbf {\bibinfo {volume}
  {3}},\ \bibinfo {pages} {817} (\bibinfo {year} {2007})}\BibitemShut {NoStop}%
\bibitem [{\citenamefont {Vlahovska}\ \emph {et~al.}(2009)\citenamefont
  {Vlahovska}, \citenamefont {Gracia}, \citenamefont {Aranda-Espinoza},\ and\
  \citenamefont {Dimova}}]{vlahovska2009electrohydrodynamic}%
  \BibitemOpen
  \bibfield  {author} {\bibinfo {author} {\bibfnamefont {P.~M.}\ \bibnamefont
  {Vlahovska}}, \bibinfo {author} {\bibfnamefont {R.~S.}\ \bibnamefont
  {Gracia}}, \bibinfo {author} {\bibfnamefont {S.}~\bibnamefont
  {Aranda-Espinoza}}, \ and\ \bibinfo {author} {\bibfnamefont {R.}~\bibnamefont
  {Dimova}},\ }\href@noop {} {\bibfield  {journal} {\bibinfo  {journal}
  {Biophysical journal}\ }\textbf {\bibinfo {volume} {96}},\ \bibinfo {pages}
  {4789} (\bibinfo {year} {2009})}\BibitemShut {NoStop}%
\bibitem [{\citenamefont {Schwalbe}\ \emph
  {et~al.}(2011{\natexlab{a}})\citenamefont {Schwalbe}, \citenamefont
  {Vlahovska},\ and\ \citenamefont {Miksis}}]{schwalbe2011}%
  \BibitemOpen
  \bibfield  {author} {\bibinfo {author} {\bibfnamefont {J.~T.}\ \bibnamefont
  {Schwalbe}}, \bibinfo {author} {\bibfnamefont {P.~M.}\ \bibnamefont
  {Vlahovska}}, \ and\ \bibinfo {author} {\bibfnamefont {M.~J.}\ \bibnamefont
  {Miksis}},\ }\href@noop {} {\bibfield  {journal} {\bibinfo  {journal}
  {Physical Review E}\ }\textbf {\bibinfo {volume} {83}},\ \bibinfo {pages}
  {046309} (\bibinfo {year} {2011}{\natexlab{a}})}\BibitemShut {NoStop}%
\bibitem [{\citenamefont {McConnell}\ \emph {et~al.}(2013)\citenamefont
  {McConnell}, \citenamefont {Miksis},\ and\ \citenamefont
  {Vlahovska}}]{mcconnell2013}%
  \BibitemOpen
  \bibfield  {author} {\bibinfo {author} {\bibfnamefont {L.~C.}\ \bibnamefont
  {McConnell}}, \bibinfo {author} {\bibfnamefont {M.~J.}\ \bibnamefont
  {Miksis}}, \ and\ \bibinfo {author} {\bibfnamefont {P.~M.}\ \bibnamefont
  {Vlahovska}},\ }\href@noop {} {\bibfield  {journal} {\bibinfo  {journal} {IMA
  Journal of Applied Mathematics}\ }\textbf {\bibinfo {volume} {78}},\ \bibinfo
  {pages} {797} (\bibinfo {year} {2013})}\BibitemShut {NoStop}%
\bibitem [{\citenamefont {Salipante}\ and\ \citenamefont
  {Vlahovska}(2014)}]{salipante2014}%
  \BibitemOpen
  \bibfield  {author} {\bibinfo {author} {\bibfnamefont {P.~F.}\ \bibnamefont
  {Salipante}}\ and\ \bibinfo {author} {\bibfnamefont {P.~M.}\ \bibnamefont
  {Vlahovska}},\ }\href@noop {} {\bibfield  {journal} {\bibinfo  {journal}
  {Soft Matter}\ }\textbf {\bibinfo {volume} {10}},\ \bibinfo {pages} {3386}
  (\bibinfo {year} {2014})}\BibitemShut {NoStop}%
\bibitem [{\citenamefont {Salipante}(2013)}]{Salipante2013thesis}%
  \BibitemOpen
  \bibfield  {author} {\bibinfo {author} {\bibfnamefont {P.}~\bibnamefont
  {Salipante}},\ }\emph {\bibinfo {title} {{Electrohydrodynamics of simple and
  complex interfaces}}},\ \href@noop {} {\bibinfo {type} {Ph.d dissertation}},\
  \bibinfo  {school} {Brown University} (\bibinfo {year} {2013})\BibitemShut
  {NoStop}%
\bibitem [{\citenamefont {Riske}\ and\ \citenamefont
  {Dimova}(2005)}]{Riske2005}%
  \BibitemOpen
  \bibfield  {author} {\bibinfo {author} {\bibfnamefont {K.~A.}\ \bibnamefont
  {Riske}}\ and\ \bibinfo {author} {\bibfnamefont {R.}~\bibnamefont {Dimova}},\
  }\href@noop {} {\bibfield  {journal} {\bibinfo  {journal} {Biophysical
  journal}\ }\textbf {\bibinfo {volume} {88}},\ \bibinfo {pages} {1143}
  (\bibinfo {year} {2005})}\BibitemShut {NoStop}%
\bibitem [{\citenamefont {Schwalbe}\ \emph
  {et~al.}(2011{\natexlab{b}})\citenamefont {Schwalbe}, \citenamefont
  {Vlahovska},\ and\ \citenamefont {Miksis}}]{schwalbe2011lipid}%
  \BibitemOpen
  \bibfield  {author} {\bibinfo {author} {\bibfnamefont {J.~T.}\ \bibnamefont
  {Schwalbe}}, \bibinfo {author} {\bibfnamefont {P.~M.}\ \bibnamefont
  {Vlahovska}}, \ and\ \bibinfo {author} {\bibfnamefont {M.~J.}\ \bibnamefont
  {Miksis}},\ }\href@noop {} {\bibfield  {journal} {\bibinfo  {journal}
  {Physics of Fluids (1994-present)}\ }\textbf {\bibinfo {volume} {23}},\
  \bibinfo {pages} {041701} (\bibinfo {year} {2011}{\natexlab{b}})}\BibitemShut
  {NoStop}%
\bibitem [{\citenamefont {Seiwert}\ \emph {et~al.}(2012)\citenamefont
  {Seiwert}, \citenamefont {Miksis},\ and\ \citenamefont
  {Vlahovska}}]{seiwert2012stability}%
  \BibitemOpen
  \bibfield  {author} {\bibinfo {author} {\bibfnamefont {J.}~\bibnamefont
  {Seiwert}}, \bibinfo {author} {\bibfnamefont {M.~J.}\ \bibnamefont {Miksis}},
  \ and\ \bibinfo {author} {\bibfnamefont {P.~M.}\ \bibnamefont {Vlahovska}},\
  }\href@noop {} {\bibfield  {journal} {\bibinfo  {journal} {Journal of Fluid
  Mechanics}\ }\textbf {\bibinfo {volume} {706}},\ \bibinfo {pages} {58}
  (\bibinfo {year} {2012})}\BibitemShut {NoStop}%
\bibitem [{\citenamefont {Kolahdouz}\ and\ \citenamefont
  {Salac}(2015{\natexlab{a}})}]{kolahdouzEHD2014}%
  \BibitemOpen
  \bibfield  {author} {\bibinfo {author} {\bibfnamefont {E.~M.}\ \bibnamefont
  {Kolahdouz}}\ and\ \bibinfo {author} {\bibfnamefont {D.}~\bibnamefont
  {Salac}},\ }\href@noop {} {\bibfield  {journal} {\bibinfo  {journal} {SIAM
  Journal on Scientific Computing}\ }\textbf {\bibinfo {volume} {Under Review}}
  (\bibinfo {year} {2015}{\natexlab{a}})}\BibitemShut {NoStop}%
\bibitem [{\citenamefont {Kolahdouz}\ and\ \citenamefont
  {Salac}(2015{\natexlab{b}})}]{kolahdouz2015}%
  \BibitemOpen
  \bibfield  {author} {\bibinfo {author} {\bibfnamefont {E.~M.}\ \bibnamefont
  {Kolahdouz}}\ and\ \bibinfo {author} {\bibfnamefont {D.}~\bibnamefont
  {Salac}},\ }\href@noop {} {\bibfield  {journal} {\bibinfo  {journal} {Applied
  Mathematics Letters}\ }\textbf {\bibinfo {volume} {39}},\ \bibinfo {pages}
  {7} (\bibinfo {year} {2015}{\natexlab{b}})}\BibitemShut {NoStop}%
\bibitem [{\citenamefont {Riske}\ and\ \citenamefont
  {Dimova}(2006)}]{Riske2006}%
  \BibitemOpen
  \bibfield  {author} {\bibinfo {author} {\bibfnamefont {K.~A.}\ \bibnamefont
  {Riske}}\ and\ \bibinfo {author} {\bibfnamefont {R.}~\bibnamefont {Dimova}},\
  }\href@noop {} {\bibfield  {journal} {\bibinfo  {journal} {Biophysical
  journal}\ }\textbf {\bibinfo {volume} {91}},\ \bibinfo {pages} {1778}
  (\bibinfo {year} {2006})}\BibitemShut {NoStop}%
\bibitem [{\citenamefont {Sadik}\ \emph {et~al.}(2011)\citenamefont {Sadik},
  \citenamefont {Li}, \citenamefont {Shan}, \citenamefont {Shreiber},\ and\
  \citenamefont {Lin}}]{sadik2011vesicle}%
  \BibitemOpen
  \bibfield  {author} {\bibinfo {author} {\bibfnamefont {M.~M.}\ \bibnamefont
  {Sadik}}, \bibinfo {author} {\bibfnamefont {J.}~\bibnamefont {Li}}, \bibinfo
  {author} {\bibfnamefont {J.~W.}\ \bibnamefont {Shan}}, \bibinfo {author}
  {\bibfnamefont {D.~I.}\ \bibnamefont {Shreiber}}, \ and\ \bibinfo {author}
  {\bibfnamefont {H.}~\bibnamefont {Lin}},\ }\href@noop {} {\bibfield
  {journal} {\bibinfo  {journal} {Physical Review E}\ }\textbf {\bibinfo
  {volume} {83}},\ \bibinfo {pages} {066316} (\bibinfo {year}
  {2011})}\BibitemShut {NoStop}%
\bibitem [{\citenamefont {Seibold}\ \emph {et~al.}(2012)\citenamefont
  {Seibold}, \citenamefont {Rosales},\ and\ \citenamefont
  {Nave}}]{Seibold2012}%
  \BibitemOpen
  \bibfield  {author} {\bibinfo {author} {\bibfnamefont {B.}~\bibnamefont
  {Seibold}}, \bibinfo {author} {\bibfnamefont {R.~R.}\ \bibnamefont
  {Rosales}}, \ and\ \bibinfo {author} {\bibfnamefont {J.-C.}\ \bibnamefont
  {Nave}},\ }\href@noop {} {\bibfield  {journal} {\bibinfo  {journal} {Discrete
  and Continuous Dynamical Systems - Series B}\ }\textbf {\bibinfo {volume}
  {17}},\ \bibinfo {pages} {1229} (\bibinfo {year} {2012})}\BibitemShut
  {NoStop}%
\bibitem [{\citenamefont {Kolahdouz}\ and\ \citenamefont
  {Salac}(2013)}]{Kolahdouz20132}%
  \BibitemOpen
  \bibfield  {author} {\bibinfo {author} {\bibfnamefont {E.~M.}\ \bibnamefont
  {Kolahdouz}}\ and\ \bibinfo {author} {\bibfnamefont {D.}~\bibnamefont
  {Salac}},\ }\href@noop {} {\bibfield  {journal} {\bibinfo  {journal} {SIAM
  Journal on Scientific Computing}\ }\textbf {\bibinfo {volume} {35}},\
  \bibinfo {pages} {231} (\bibinfo {year} {2013})}\BibitemShut {NoStop}%
\bibitem [{\citenamefont {Smereka}(2003)}]{Smereka2003}%
  \BibitemOpen
  \bibfield  {author} {\bibinfo {author} {\bibfnamefont {P.}~\bibnamefont
  {Smereka}},\ }\href@noop {} {\bibfield  {journal} {\bibinfo  {journal}
  {Journal of Scientific Computing}\ }\textbf {\bibinfo {volume} {19}},\
  \bibinfo {pages} {439} (\bibinfo {year} {2003})}\BibitemShut {NoStop}%
\end{thebibliography}
\end{document}